\documentclass[]{aa}
\usepackage{natbib}
\usepackage{graphicx}
\usepackage{amsmath}
\begin{document}
\title{Does the butterfly diagram indicate a 
       solar flux-transport dynamo?}
\titlerunning{Butterfly diagram and dynamo models}
\author{M. Sch\"ussler \and D. Schmitt}
\institute{Max-Planck-Institut f\"ur Sonnensystemforschung\footnote[1]{formerly: Max-Planck-Institut f\"ur Aeronomie},
Max-Planck-Str. 2, 37191 Katlenburg-Lindau, Germany \\
\email{schuessler@linmpi.mpg.de,schmitt@linmpi.mpg.de}}
\date{Received; accepted}
\abstract{
We address the question whether the properties of the observed
latitude-time diagram of sunspot occurence (the butterfly diagram)
provide evidence for the operation of a flux-transport
dynamo, which explains the migration of the sunspot zones and the
period of the solar cycle in terms of a deep equatorward
meridional flow.  We show that the properties of the butterfly diagram
are equally well reproduced by a conventional dynamo model with
migrating dynamo waves, but without transport of magnetic flux by a
flow. These properties seem to be generic for an oscillatory and
migratory field of dipole parity and thus do not permit an
observational distinction between different dynamo approaches.
\keywords{MHD --- Sun: activity --- Sun: magnetic fields --- 
          Sun: interior}}
\maketitle

\section{Introduction}
In a recent paper, \citet[][henceforth referred to as
HNWR]{Hathaway:etal:2003} presented an analysis of the latitude-time
diagram of sunspot observations (commonly called butterfly diagram) and
suggested that their results provide ``strong observational evidence
that a deep meridional flow toward the equator is driving the sunspot
cycle''. This refers to the so-called flux-transport dynamo models,
which attribute the equatorward drift of the sunspot zone in the course
of the 11-year solar activity cycle to the transport of toroidal
magnetic flux towards low heliographic latitudes by an equatorward
meridional flow near the bottom of the convection zone, thought to be
the return flow of the observed poleward flow in the upper part of the
convection zone and at the solar surface. Such physical transport of
magnetic flux is not used in the more `traditional' type of dynamo
models, which explain the equatorward drift by a latitudinally
propagating dynamo wave and thus do not require a material flow
\citep[for a recent comprehensive review of solar dynamo theory,
see][]{Ossendrijver:2003}. In this paper we show that the properties of
the butterfly diagram analysed by HNWR (drift velocity of the sunspot
zone as a function of latitude and its relation to cycle length and
amplitude) are well consistent with a dynamo-wave model without
meridional flow.


\section{Dynamo model and results}
We use a dynamo-wave model without meridional flow to obtain a synthetic
butterfly diagram whose basic features are consistent with the solar
case, i.e., the magnetic field is concentrated in low latitudes, is
antisymmetric with respect to the equator, and reverses polarity from
one cycle to the next. We then analyse the properties of the synthetic
butterfly diagram in an analogous way as HNWR did with the observed
data. It is not our intention here to advocate specific dynamo concepts or
models, our sole goal is to clarify whether the observations in
fact exclude a dynamo-wave model. Therefore, we do not aim at a
completely realistic and detailed model for the solar cycle and thus
restrict ourselves to a simple quasi-1D $\alpha\Omega$-dynamo model
\citep{Schmitt:Schuessler:1989,Hoyng:etal:1994} driven by radial
differential rotation and by an $\alpha$-effect due to buoyancy
instability of the toroidal magnetic field
\citep{Schmitt:1987,Ferriz-Mas:etal:1994}. For simplicity, we assume a
constant radial gradient of rotation and a cosine-shaped profile of the
$\alpha$-effect extending from 0 to $60 \deg$ latitude. The dynamo
amplitude is limited by a nonlinearity mimicking the buoyant loss of
magnetic flux from the dynamo region at the bottom of the convection
zone. We allow for random fluctuations of the $\alpha$-effect in order
to simulate the irregularity of the solar cycle.

\begin{figure*}
 \begin{center}
  \includegraphics[width=\hsize]{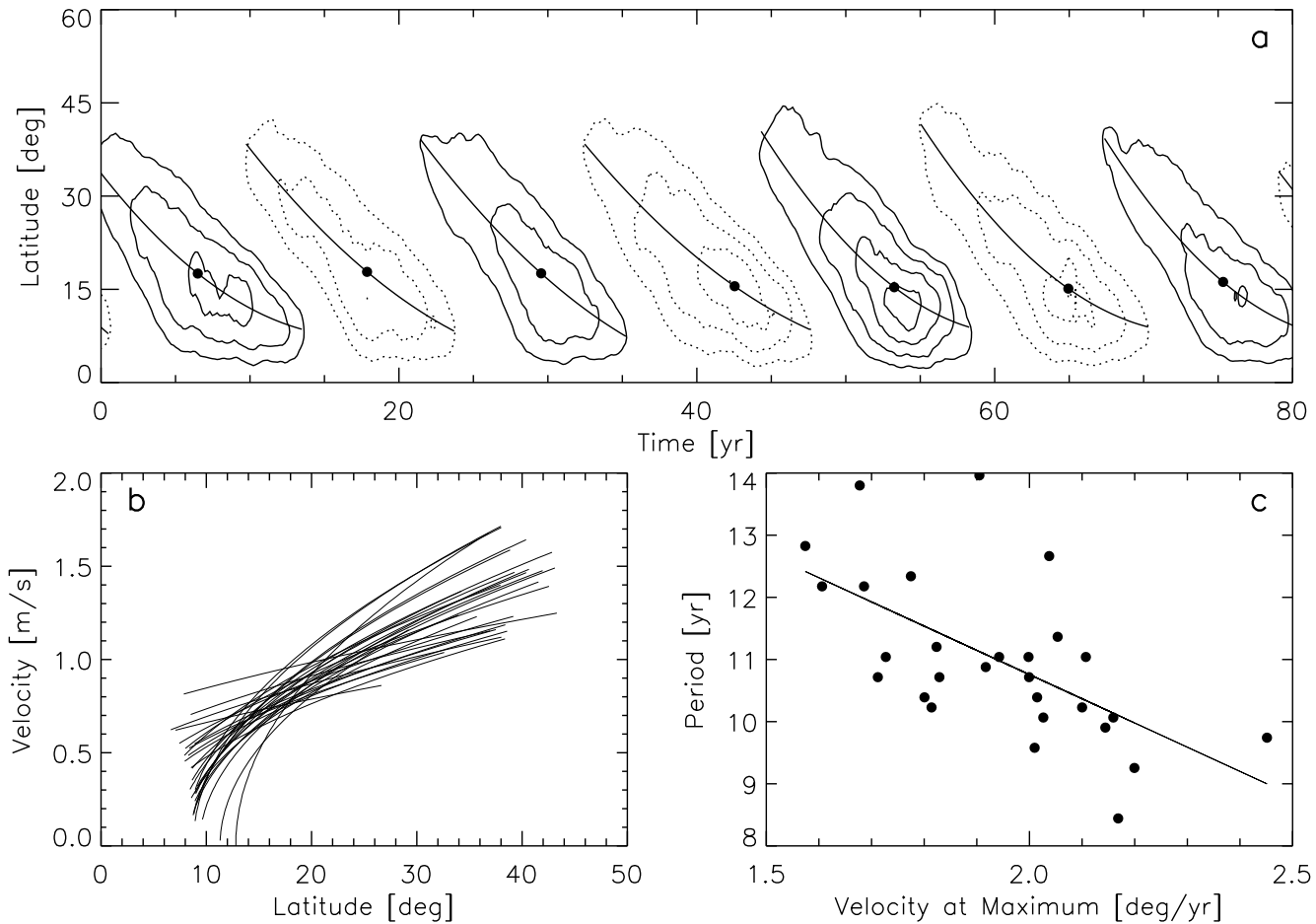}
 \end{center}
 \caption{Results obtained with
      an $\alpha\Omega$ dynamo model providing latitudinally propagating
      dynamo waves. {\bf a}: latitude-time diagram (butterfly diagram)
      of the toroidal magnetic field. Full lines indicate positive,
      dashed lines negative values. The butterfly wings on the southern
      hemisphere (not shown) are the opposite-polarity mirror images of
      the wings shown in the graph. The time unit (diffusion time) has
      been adjusted to obtain an average cycle period of 11 years for
      the analyzed time series of 28 cycles. The curved diagonal lines
      are the drift curves of the butterfly wings with black dots
      indicating the times of maximum magnetic energy. The variability
      of the individual cycles results from a stochastic variation of
      the dynamo excitation.  {\bf b}: latitudinal drift velocity as a
      function of latitude for 28 simulated  cycles. The
      deceleration of the drift near the equator is obvious. {\bf c}:
      cycle period (time between consecutive minima of the energy in the
      toroidal magnetic field) vs. drift velocity at the corresponding
      cycle maximum. The dots indicate the 28 simulated cycles
      analyzed. The line represents a least-square linear fit.}
 \label{fig1}
\end{figure*}

It is customary to take the strength of the toroidal field, $B$, in the
dynamo region as a proxy for magnetic flux eruption and compare
latitude-time diagrams of $B$ with the solar butterfly diagram.
Fig.~\ref{fig1}a shows a section from such a diagram produced by our
dynamo model. The contour lines correspond to $\pm 0.3,0.5,0.7,0.9$
times $B_{\rm max}$, the maximum toroidal field strength reached during
the time considered. The drift curves for the individual butterfly wings
are shown on the same graph. These have been determined in analogy to
the procedure used by HNWR. We first define the centroid positions of
the wings for each time step as the median position of the latitude
profile of $\vert B\vert$ between the outermost contour lines
(corresponding to $B=\pm 0.3B_{\rm max}$, taken as the threshold for the
onset of sunspot activity). A quadratic function is then fitted to the
centroid positions for each butterfly wing to obtain the drift
curves. The times of maximum energy of the toroidal magnetic field
(`sunspot maxima') are indicated by black dots on the drift curves.

Figure~\ref{fig1}b shows the profiles of the drift velocity (corresponding
to the slope of the drift curves) as a function of latitude for a sample
of 28 consecutive butterfly wings, including those shown in
Fig.~\ref{fig1}a.  Comparison with Fig.~3 of HNWR reveals a striking
similarity of both figures. In particular, the deceleration towards the
equator, which has been taken by HNWR as evidence for the magnetic field
being carried by a flow turning upward near the equator, is perfectly
reproduced by a dynamo wave in the absence of any meridional flow. The
scatter among the curves is caused by the random variation of the dynamo
excitation ($\alpha$-effect) in our model.

A similar agreement between the dynamo-wave model and observation is
found concerning the relation between cycle period and drift velocity at
activity maximum. Fig.~\ref{fig1}c shows a clear anticorrelation between
these quantities for the 28 simulated cycles (dots); the correlation
coefficient is $-0.61$ with a $t-$value of 3.9, corresponding to a
confidence level of 99.9\%. Again, this result is in good qualitative
agreement with the data analysis of HNWR (their Fig.~4). Our dynamo-wave
model shows that the anticorrelation between drift velocity and cycle
period can be reproduced without any meridional flow that sets the
period. In our case, the variation of the cycle period is determined by
the stochastically varying dynamo excitation: larger $\alpha$-effect
leads to shorter period, and vice versa. 

Note that there is no need for fine-tuning of the model parameters in
order to arrive at results that compare well with the observed
properties.  Our results are robust with respect to variations of our
model parameters as long as the butterfly diagrams remain basically
solar-like, i.e., antisymmetric, migrating equatorwards, and
concentrated towards low latitudes. Otherwise, a comparison with
observation would not be meaningful anyway.

\section{Discussion}
We have demonstrated that the properties of the observed butterfly
diagram are consistent with dynamo-wave models as well as with
flux-transport dynamos. In fact, these properties (slowing of the drift
near the equator and anticorrelation between cycle length and drift
velocity) appear to be generic features of a toroidal field of dipole
parity (antisymmetric with respect to the equator) performing a periodic
equatorward drift. Since such a field vanishes at the equator, the
drifting toroidal field patterns in both hemispheres have to stop there.
Diffusion then leads to a smooth decrease of the drift velocity when
approaching the equator, independent of whether the drift is caused by a
material motion (meridional flow) or by a dynamo wave. The
anticorrelation of the drift velocity of the butterfly wings with the
cycle period is almost trivial and largely independent of the physical
ingredients of dynamo models: when the period is shorter, the butterfly
wings traverse the latitude range of activity within a shorter time and
thus travel with a larger speed.

The relationship between drift rate and cycle amplitudes also discussed
by HNWR is equivalent to the well-known (weak) anticorrelation between
cycle length and amplitude dating back to the days of Rudolf
\citet{Wolf:1861}. The authors mention as another factor in favor of
flux-transport dynamo models that there is a stronger anticorrelation
between the cycle length and the amplitude of the {\em next}
cycle. However, for the time interval analyzed by HNWR, the strongest
anticorrelation in fact appears between the length of cycle $n$ and the
amplitude of cycle $n+3$ \citep{Solanki:etal:2002b}, which is difficult
to explain in terms of {\em any} existing dynamo model.

We do not claim that our simple dynamo-wave model represents a
realistic description of the solar conditions.  In fact, the sign
reversal of the radial differential rotation in higher latitudes is
not included and could possibly (depending on the $\alpha$-effect, see
below) lead to a poleward migrating branch of the dynamo wave.  On the
other hand, such a branch would probably stay unobservable at the
solar surface since the magnetic buoyancy instability at high
latitudes sets in only for significantly larger field strength than
near the equator
\citep{Schuessler:etal:1994,Ferriz-Mas:Schuessler:1995}, so that no
large-scale magnetic flux would emerge in the polar regions
anyway. The concentration of the $\alpha$-effect to low latitudes and
the choice of its sign (which determines the propagation direction of
the dynamo wave) may seem arbitrary, but note that the buoyancy
instability of toroidal magnetic field yields an $\alpha$-effect with
a similar low-latitude profile and sign
\citep{Schmitt:1987,Schmitt:2003}. There is even a mid-latitude sign
change of the $\alpha$-effect in that model, which, together with the
sign change of radial differential rotation, could again lead to a
uniformly equatorward propagating dynamo wave.  Anyway, the degree of
arbitrariness in our model and parametrization is certainly not larger
than that of flux-transport dynamo models, which have to specify the
unknown properties of the deep meridional flow (depth extension, flow
geometry and speed) in addition to the profile of the
$\alpha$-effect. No existing dynamo model does actually {\em predict}
properties like the cycle length or the latitude extension of the
butterfly wings.

\section{Conclusion}
Results of a simple dynamo model show that the drift velocity of the
sunspot zone as a function of latitude and its relation to cycle length
and amplitude can be reproduced by a migrating dynamo wave.  This casts
doubt upon the suggestion of \citet{Hathaway:etal:2003} that these
properties provide observational evidence for a flux-transport dynamo
based upon an equatorward meridional flow in the deep convection zone
and that this flow sets the cycle period. In fact, these properties of
the butterfly diagram seem to be generic to a field of dipolar parity
with equatorward drifting, opposite-polarity branches of toroidal field.

Our result that the butterfly diagram does not permit an observational
distinction between dynamo-wave and flux-transport models does not
lessen the appeal of the flux-transport dynamo concept.  Indeed, the
poleward meridional flow in the outer parts of the convection
zone is an observed fact and clearly mass conservation requires an
equatorward return flow somewhere below. However, whether the
properties of this flow meet the requirements of flux-transport dynamo
models can only be clarified by helioseismic measurements of the
meridional flow throughout the whole convection zone and over a time
period of the order of the solar cycle \citep[cf.][]{Haber:etal:2002}.

\bibliography{0341.bbl}

\end{document}